\documentclass[twocolumn,pre,floats,aps,amsmath,amssymb]{revtex4}
\usepackage{graphicx}
\usepackage{bm}
\usepackage{epstopdf}

\renewcommand\appendix{\par
\setcounter{section}{0}%
\setcounter{subsection}{0}%
\setcounter{table}{0}
\setcounter{figure}{0}
\gdef\thetable{\Alph{table}}
\gdef\thefigure{\Alph{figure}}
\section*{Appendix: theory of mode structure in perpendicularly magnetized discs}
\label{sec:appendix}
\gdef\thesection{\Alph{section}}
\setcounter{section}{1}}

\begin{document}

\title{Standing spin-wave mode structure and linewidth in partially disordered perpendicular magnetized sub-micron Permalloy disc arrays}
\author{N. Ross, and M. Kostylev}
\affiliation{School of Physics, University of Western Australia, Crawley, WA, Australia}
\author{R. L. Stamps}
\affiliation{SUPA School of Physics and Astronomy, University of Glasgow, Glasgow G12 8QQ, United Kingdom}
\affiliation{School of Physics, University of Western Australia, Crawley, WA, Australia}
\date{\today}

\begin{abstract}
Standing spin wave mode frequencies and linewidths in partially disordered perpendicular magnetized arrays of sub-micron Permalloy discs are measured using broadband ferromagnetic resonance and compared to analytical results from a single, isolated disc. The measured mode structure qualitatively reproduces the structure expected from the theory. Fitted demagnetizing parameters decrease with increasing array disorder. The frequency difference between the first and second radial modes is found to be higher in the measured array systems than predicted by theory for an isolated disc. The relative frequencies between successive spin wave modes are unaffected by reduction of the long-range ordering of discs in the array. An increase in standing spin wave resonance linewidth at low applied magnetic fields is observed and grows more severe with increased array disorder.


\end{abstract}

\maketitle

\section{Introduction}
\label{sec:intro}

Understanding the high-frequency dynamics of sub-micron diameter, nanometer thickness magnetic discs is important for potential applications in data storage \cite{TerrisandThomson2005,Thomsonetal2006,Shawetal2007,Hellwigetal2007,Shawetal2008} and spintronics \cite{Kiselevetal2003,Deacetal2008} technologies. There have been a number of recent studies concerned with spin wave mode structure \cite{Jorzicketal1999,GuslienkoandSlavin2000,JungWatetal2002,Jungetal2002,PolitiandPini2002,Rivkinetal2004,Gubbiottietal2006,Giovanninietal2007,Rivkinetal2007PRB,Rivkinetal2007JMMM,Nevirkovetsetal2008,Semenovaetal2013,Tacchietal2010} and linewidths \cite{Schneideretal2007,Shawetal2009,Rivkinetal2009,Rossetal2010,Kakazeietal2008} in the in-plane magnetized configuration of such dipole coupled arrays. There have also been studies conducted for disc arrays in the perpendicularly magnetized state, with characterization achieved by cavity ferromagnetic resonance \cite{Kakazeietal2004,Kostylevetal2008} (FMR) and magnetic resonance force microscopy \cite{Mewesetal2006}. More recently, broadband FMR has been used to study highly ordered arrays of micrometer diameter, nanometer thickness discs \cite{Casteletal2012} and nanometer diameter antidots \cite{Balietal2012}. The large range of resonant frequencies and fields available by the broadband FMR technique has not yet been utilised to study the mode structure and linewidth in closely packed arrays of dipole-coupled magnetic nanostructures in the perpendicularly saturated state.

In this study, broadband FMR was used to study the perpendicularly magnetized spin wave mode structure and linewidth of a series of four disc array samples with varying degrees of array ordering, over a wide range of excitation frequencies. These dipole-coupled arrays have been previously studied in the tangentially magnetized state, and details of their production and characterization can be found in Ref. \cite{Rossetal2010}. Each of the four array samples consisted of a locally trigonal array of Permalloy discs, with each array distinguished from the others by a different degree of long-range ordering. The long-range ordering was quantified with a parameter $\phi'$, the average amount of variation in the lattice angle in degrees per millimeter.

\begin{figure*}[hbt]
\begin{equation}
\omega_m^2 = \gamma^2 (H - 4 \pi M_S N_m + \frac{2 A k_m^2}{M_S}) (H - 4 \pi M_S N_m + \frac{2 A k_m^2}{M_S} + 4 \pi M_S (1- \frac{1-e^{-k_m L}}{k_m L}))\label{eq:omega_squared},
\end{equation}
\end{figure*}

\section{Experiment}
\label{sec:experiment}

\begin{table}[th]
\begin{center}
\begin{tabular}{ | c || c   c   c   | }
    \hline
    Film & d [nm] & $\phi' [{}^{\circ}~\mathrm{mm^{-1}}]$   & $4 \pi M_S$ [kOe] \\ \hline \hline
   f3c & - & - & 8.49 \\
   f3b & $695 \pm 28$ & $6.0 \pm 0.8$ & -  \\ 
   f3a & $703 \pm 37$ & $9.4 \pm 1.1$ & -   \\
   f2c & - & - & 8.69 \\
   f2a & $697 \pm 31$ & $11.3 \pm 1.7$ & -   \\
   f1c & - & - & 8.85 \\
   f1a & $699 \pm 28$ & $19.9 \pm 2.1$ & -   \\
    \hline
\end{tabular}
\end{center}
\caption{Table showing average disc diameter, $d$, array angular variation per unit length $\phi'$, and saturation magnetization $4 \pi M_S$ for the samples used in this study. All discs and films had thickness $ 27 \pm 3$ nm.}
\label{tab:samples}
\end{table}

\begin{figure}[ht]
\includegraphics[width=8.5cm]{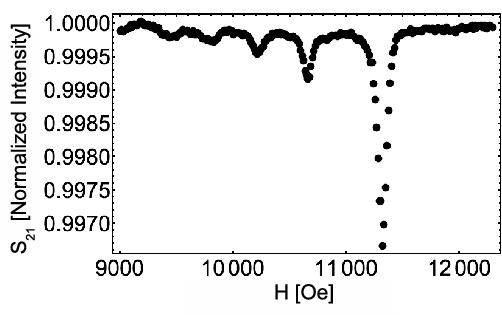}
\caption{Plot of normalized transmission parameter ${S}_{21}$ vs applied out-of-plane magnetic field $H$ for sample f2a, at an excitation frequency of 10 GHz, as measured by Broadband FMR. Five radial modes are resolved.}
\label{fig:f2a_mode_structure}
\end{figure}

The structural characteristics of each samples are listed in Table \ref{tab:samples}. Samples f3c, f2c, and f1c were continuous film sections cut from the parent film from which f3a and f3b, f2a, and f1a respectively were patterned. The samples were placed face-down on an 8 mil microstrip waveguide connected to a two-port vector network analyzer and magnetized perpendicular-to-plane with respect to the substrate. The broadband FMR measurement was performed at frequencies between 6 and 17 GHz, in intervals of one GHz. The microwave transmission parameter $S_{21}$ was measured as the applied magnetic field was swept through the experimentally available range, in analogy to the cavity FMR experiment. Negligible reflections allowed $S_{11}$ to be ignored \cite{Couniletal2004}. An example of the spin wave spectra obtained is shown in Figure \ref{fig:f2a_mode_structure}. This spectrum of modes is in qualitative agreement with those measured previously \cite{Kakazeietal2004,Kostylevetal2008,Mewesetal2006,Casteletal2012}, and arises directly from the cylindrical symmetry of the discs in the axially magnetized configuration, as shown in the appendix. The frequency of these modes can be expressed as in Eq. \ref{eq:omega_squared}, where $k_{m}$ is the in-plane wave vector, order $m=1,2,3...$, $\gamma$ the gyromagnetic ratio, $H$ the applied magnetic field, $M_S$ the saturation magnetizatio, $A$ the exchange stiffness constant, and $L$ the disc thickness.


Previous investigation of these array samples \cite{Rossetal2010} showed no evidence of significant material or structural differences between the four samples. Specifically, atomic force microscopy images showed that all four samples had smooth disc boundaries, and there was no significant evidence in the SQuID-measured in-plane hysteresis loops of structures which do not support compensated magnetic vortices at remanence. With the exception of small differences in saturation magnetizations $M_S$ and parent film linewidths $\Delta H$, the only parameter known to vary significantly between the arrays was the degree of long-range array ordering, $\phi'$.


\section{Results}
\label{sec:results}

\begin{figure}[ht]
\includegraphics[width=8.5cm]{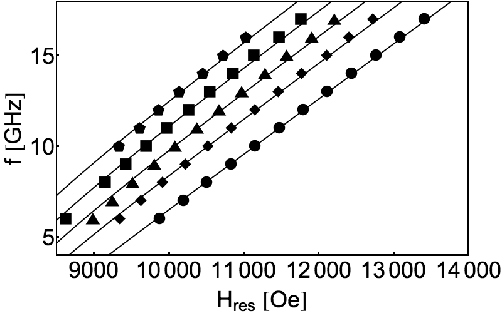}
\caption{Plot of frequency $f$ vs out-of-plane magnetic resonance field ${H}_\mathrm{res}$ for the first five radial modes of sample f3b (circles: $m=1$; diamonds: $m=2$; triangles: $m=3$; squares: $m=4$; pentagons: $m=5$), as measured by Broadband FMR. Solid lines are fits to Equation \ref{eq:omega_squared}. These data are qualitatively representative of those data obtained from all of the samples.}
\label{fig:f_vs_H}
\end{figure}

The range of applied magnetic fields available for the field sweep was defined at its lower end by the minimum field expected to be necessary to saturate the magnetization of the discs out of the substrate film plane, $H \simeq 4 \pi M_S$, and at its upper end by the maximum field attainable with the available electromagnet, $H=14$ kOe. A representative plot of the resonance frequencies $f$ of the first five radial modes against the resonance field $H_{res}$ in this field range is shown in Figure \ref{fig:f_vs_H}. The solid lines are the fits for the data to Equation \ref{eq:omega_squared}, having left $N_m$ as the free fitting parameter in each case and otherwise using measured film characteristics.

The fitted values of $N_m$ are tabulated in Table \ref{tab:Nm}, along with the values of $N_m$ calculated for a theoretical isolated `disc', 700 nm diameter, 27 nm thickness. The uncertainties in the $N_m$ values were calculated by propagating the uncertainties in the disc radius $R$ and thickness $L$ through the fit. There was a slight decrease in demagnetising parameter $N_m$ with increasing array disorder and increasing mode number. The changes in $N_m$ across the range of $\phi'$ available were small enough to be comparable with the uncertainties. However, $N_m$ values for an isolated disc differ from the $N=1$ for a continuous film by less than 0.07. The change in $N_m$ with reduced array ordering cannot be expected to be higher than this difference value.

\begin{table*}[th]
\begin{center}
\begin{tabular}{ | c | c ||  c   c   c   c   c  | }
    \hline
    Film & $\phi' [{}^{\circ}~\mathrm{mm^{-1}}]$   & $\mathrm{{N}_1}~\times~10$ & $\mathrm{{N}_2}~\times~10$ & $\mathrm{{N}_3}~\times~10$ & $\mathrm{{N}_4}~\times~10$ & $\mathrm{{N}_5}~\times~10$ \\ \hline \hline
    'disc' & - & $9.51$ & $9.40$ & $9.35$ & $9.33$ & $9.31$ \\
    f3b & $6.0 \pm 0.8$ & $9.67 \pm 0.05$ & $9.44 \pm 0.08$ & $9.38 \pm 0.10$ & $9.37 \pm 0.12$ & $9.39 \pm 0.15$ \\ 
   f3a &  $9.4 \pm 1.1$ & $9.69 \pm 0.08$ & $9.43 \pm 0.14$ & $9.37 \pm 0.18$ & $9.32 \pm 0.21$ & $9.37 \pm 0.23$  \\
  f2a & $11.3 \pm 1.7$ & $9.65 \pm 0.04$ & $9.38 \pm 0.08$ & $9.32 \pm 0.10$ & $9.28 \pm 0.12$ & $9.34 \pm 0.14$  \\
   f1a & $19.9 \pm 2.1$ & $9.49 \pm 0.04$ & $9.21 \pm 0.08$ & $9.12 \pm 0.10$ & $9.08 \pm 0.12$ & $9.07 \pm 0.14$  \\
    \hline
\end{tabular}
\end{center}
\caption{Table of demagnetizing factors ${N}_m$ as calculated from Equation \ref{eq:Nmeqn} for a disc 700 nm diameter, 27 nm thickness (denoted 'disc'), and for the disc array samples f3b, f3a, f2a, and f1a, as extracted by fitting equation \ref{eq:omega_squared} to the data in Figure \ref{fig:f_vs_H}.}
\label{tab:Nm}
\end{table*}

In addition to changes with $m$ and $\phi'$, there was a dramatic difference in the contrast between analytical and measured $N_1$ and $N_2$, the demagnetizing parameters for the first and second radial modes. Displayed across the all four samples was the phenomenon  that the difference between $N_1$ and the higher order demagnetizing factors $N_{2,3,4,5}$ was larger than for the isolated analytical disc. This difference is most easily observed when framed as an average frequency difference between the fit lines of Figure \ref{fig:f_vs_H}. The average frequency differences between successive nodes are tabulated in Table \ref{tab:fdiff}, and show that the frequency difference for the first two modes, $\overline{\mathrm{f}_2-\mathrm{f}_1}$ is larger in all cases than for the theoretical, isolated disc.

\begin{table}[th]
\begin{center}
\begin{tabular}{ | c ||  c   c   c   c  | }
    \hline
    Film &  $\overline{\mathrm{f}_2-\mathrm{f}_1}$ & $\overline{\mathrm{f}_3-\mathrm{f}_2}$ & $\overline{\mathrm{f}_4-\mathrm{f}_3}$ & $\overline{\mathrm{f}_5-\mathrm{f}_4}$  \\ \hline \hline
    'disc' & 1.59 & 1.42 & 1.42 & 1.48  \\
    f3b &  $1.93 \pm 0.09$ & $1.44 \pm 0.06$ & $1.36 \pm 0.05$ & $1.40 \pm 0.08$  \\ 
   f3a &   $2.00 \pm 0.17$ & $1.44 \pm 0.09$ & $1.49 \pm 0.09$ & $1.39 \pm 0.07$   \\
  f2a &  $2.01 \pm 0.09$ & $1.44 \pm 0.06$ & $1.47 \pm 0.06$ & $1.26 \pm 0.04$   \\
   f1a & $2.03 \pm 0.09$ & $1.55 \pm 0.06$ & $1.47 \pm 0.06$ & $1.46 \pm 0.05$   \\
    \hline
\end{tabular}
\end{center}
\caption{Table of average frequency differences in GHz between fits to Equation \ref{eq:omega_squared} of successive modes, $\overline{\mathrm{f}_{m+1}-\mathrm{f}_m}=1/(H_\mathrm{max}-H_\mathrm{min}) \int_{H_\mathrm{min}}^{H_\mathrm{max}} (\omega_{m+1}-\omega_m)/(2\pi) dH $, on the field interval [10000-12000] Oe for a disc of 700 nm diameter, 27 nm thickness (denoted `disc'), with a saturation magnetization $M_S$ and spectroscopic splitting factor $g$ identical to film f3c, and for disc array samples f3b, f3a, f2a, and f1a.}
\label{tab:fdiff}
\end{table}

\begin{figure*}[ht]
\includegraphics[width=18cm]{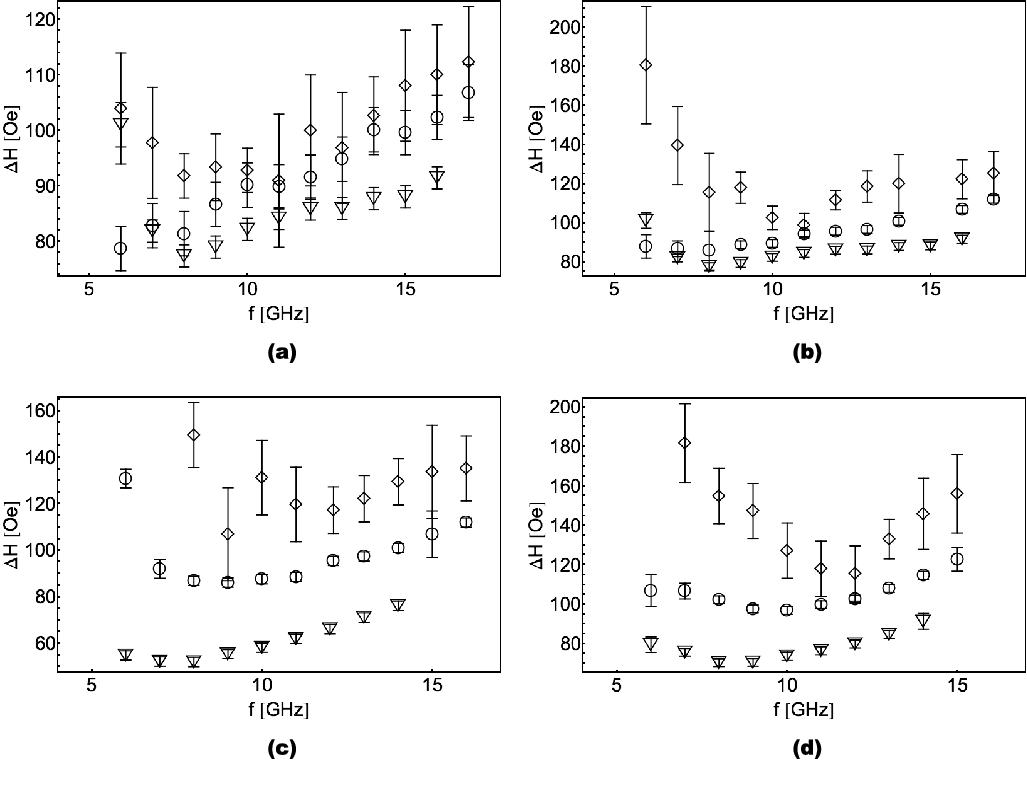}
\caption{Plot of FMR linewidth $\Delta H$ vs frequency $f$ for samples (a) f3b, (b) f3a, (c) f2a, and (d) f1a, for applied field in the direction of the film perpendicular, as measured by Broadband FMR. The first (circles) and second (diamonds) radial modes are shown alongside the data from the corresponding parent continuous films (inverted triangles).}
\label{fig:deltaH_vs_f}
\end{figure*}

The very small amplitudes of modes $m=2$ and the small magnetic background signal precluded the meaningful extraction of linewidths from those modes. Plots of the field linewidths $\Delta H$ extracted from Lorentzian fits to the first two radial modes of all four samples are shown in Figure \ref{fig:deltaH_vs_f}, alongside the linewidth data from the corresponding parent continuous films.

In previous studies of spin wave mode broadening in thin films \cite{Kalarickaletal2006} and patterned structures \cite{Shawetal2009,Rossetal2010,Casteletal2012}, the effects of intrinsic and extrinsic damping have been separated by fitting the data with the equation:

\begin{equation}
\Delta H = \Delta H_0 + \frac{4 \pi \alpha}{\gamma \mu_0}f \label{eq:deltaH}
\end{equation}

$\alpha$ is the intrinsic damping parameter in the Landau-Lifshitz-Gilbert equation \cite{LandauandLifshitz1935,Gilbert1956}, the `viscous' damping of energy to the lattice \cite{Lenzetal2006}, and $\Delta H_0$ is a term representing inhomogeneous broadening. However, $\Delta H$ does not increase with frequency in the affine fashion expected from Equation \ref{eq:deltaH}, even in the case of the parent continuous films. Instead, at low frequency values the linewidths are very broad, decreasing to some minimum, then increasing with increasing frequency in an approximately linear fashion from some onset frequency, or equivalently from the resonance field corresponding to that frequency. The linewidth in the $m=2$ mode is always larger than in the $m=1$ mode. This effect is more severe for films with higher disorder parameter $\phi'$.

\section{Discussion}
\label{sec:discussion}

As the disc radii, film thickness, and processing conditions were very similar between all of the array samples, the slight decrease in demagnetizing parameter $N_m$ with increasing array disorder $\phi'$ may be attributable to a reduction in average dipole coupling strength due to lowered symmetry and/or a slight reduction in neighbour density associated with increased array disorder. The change in $N_m$ across the range of films was between 0.018 for $m=1$ and 0.032 for $m=5$, or up to half the $N_m$ value difference between a continuous thin film and an isolated disc. Given that the same samples showed no correlation between the demagnetising parameter and $\phi'$ in the in-plane magnetized configuration \cite{Rossetal2010}, the argument from array packing density is less persuasive than the conjecture that the reduced long-range ordering impacted on the static demagnetizing field.


The difference $N_1-N_2$ or equivalently the average frequency separation between first and second modes $\overline{\mathrm{f}_2-\mathrm{f}_1}$ was larger than expected from Eq. \ref{eq:omega_squared} for an uncoupled disc. Such an effect was not observed in studies of resonance frequency of either square arrays of lower magnetization nickel discs \cite{Kakazeietal2004} or Permalloy discs with large disc-to-disc spacings \cite{Casteletal2012}. Furthermore, differences between subsequent modes after the first, for example $\overline{\mathrm{f}_3-\mathrm{f}_2}$ et cetera, were unremarkable in comparison to the analytical treatment. For a given sample, all of the radial modes were the result of the same out-of-plane static magnetization configuration, effectively ruling out static dipole coupling between the equilibrium magnetic moments of the discs as the cause of the difference in demagnetizing factors. On the other hand, the dynamic stray field of the first radial mode must be stronger than for the other modes, since it alone has no nodes across the diameter of the disc (see Figure \ref{fig:mode_structure} in the Appendix). Tacchi \emph{et al} have observed dynamic dipole coupling in travelling Bloch waves in closely packed square element arrays of comparable array element separation \cite{Tacchietal2010}, but only in the fundamental and 1DE modes. The higher order modes did not display the same coupling because of their reduced stray fields. The difference between $N_1-N_2$ and successive mode differences $N_2-N_3$, etc, in the disc arrays studied here is therefore interpreted as a dynamic dipole coupling between elements in the array which is larger for the $m=1$ mode than for subsequent modes.



The non-linear behaviour of the linewidth with decreasing frequency and applied magnetic field seen in Figure \ref{fig:deltaH_vs_f} was either not present or present in a more subtle manner in the largely dipole-uncoupled sparse square array system of Castel \emph{et al} \cite{Casteletal2012}. In square arrays of anti-dots, applied fields near the bulk saturation magnetization of the film allow canting of static magnetization vectors, leading to non-vanishing ellipticity of moment precession \cite{Balietal2012}. In the exchange coupled antidot arrays, this manifests as a measured decrease in resonance frequency below 10 GHz. For the study presented here, no significant deviation of the resonance frequency is observable in the frequency range in which the linewidth broadens anomalously. However, the linewidth broadening is more severe for less ordered arrays, and also is stronger in the $m=2$ mode than the $m=1$ mode. It is therefore likely the result of two static dipole effects: both the canting of the magnetization of a single dot away from the film perpendicular near saturation, and the distribution of this canting due to static dipole coupling of dots across the imperfectly ordered array.


\section{Conclusion}
\label{sec:conclusion}

The standing spin wave mode structure and linewidth broadening in a series of closely packed trigonal sub-micron diameter disc arrays in the perpendicularly magnetized state was investigated using broaband ferromagnetic resonance. Comparison of measurements to theory over a wide range of frequencies allowed deviations of the mode structure from that of an isolated disc to be identified. These deviations revealed the importance of array ordering to the absolute size of the demagnetizing factors. Deviations from the expected frequency differences between the first and second modes suggested that dynamic dipole coupling between discs was important to the mode structure. The relative values of demagnetizing fields for successive modes were essentially unaffected by degradation of the array symmetry.

Anomalously large linewidth broadenings were observed close to the out-of-plane demagnetizing fields of the arrays. The increased severity of this broadening in the second radial mode compared to the first suggests that the broadening is the result of the static magnetization configuration of the discs. The effect was more severe in less ordered arrays, suggesting that the increase in linewidth was the result of both the canting of local magnetic moments, and of the distribution of the canting across the partially disordered array.

\begin{acknowledgments}
This work was supported by the Australian Research Council. N. Ross was supported by a University of Western Australia Hackett Postgraduate Scholarship.

\end{acknowledgments}

\appendix
The theory for FMR modes in isolated, perpendicularly magnetized discs was first applied to sub-micron discs by Kakazei \emph{et al} in Ref \cite{Kakazeietal2004}, and is based on the dipole-exchange theory of spin wave spectra in unrestricted in-plane magnetic films \cite{KalinkosandSlavin1986,Kalinkosetal1990} and a method of the calculation of demagnetizing fields in nonellipsoidal bodies \cite{JosephandSchlomann1965}. This model does not take into account the interaction between elements in a closely packed array and therefore does not take into account the effects of disorder in that array.

In this theory, the finite radius $R$ of the disc is considered to allow only discrete values of the in-plane wave vector: $k \rightarrow k_m,~m = 1, 2, 3...$. The nonellipsoidal geometry of the disc means that the demagnetizing field inside the dot is inhomogeneous, with the internal bias field becoming a function of the disc radius $\rho$. The strong dipolar pinning at the disc edges produces dipolar eigenmodes with zeroth-order Bessel function profiles:
\begin{align}
\mu_m(\rho) &= J_0 (k_m \rho) \label{eq:mu_of_rho}
\end{align}
The standing mode profiles of the first five modes of such a disc geometry are shown in Figure \ref{fig:mode_structure}. These profiles have been confirmed experimentally by Mewes \emph{et al} in square arrays of sub-micro diameter discs using MRFM \cite{Mewesetal2006}. These mode profiles correspond in principle to the peaks observed in Figure \ref{fig:f2a_mode_structure}, with the largest amplitude, highest-field peak the $m=1$ mode. Assuming that there is no significant out-of-plane film anisotropy, the frequencies of these modes are given by Equation \ref{eq:omega_squared},with $N_m$ determined as in Reference \cite{Guslienkoetal2003}:

\begin{figure}[ht]
\includegraphics[width=8.5cm]{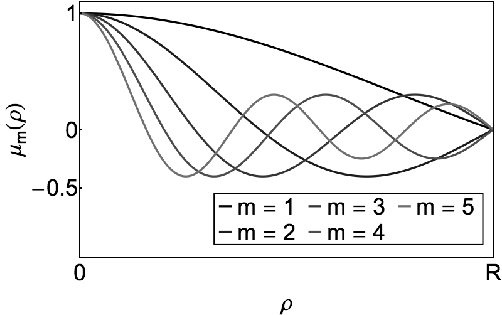}
\caption{Plot of the theoretical spin wave eigenmodes ${\mu}_m(\rho)$ for the first five modes for a perfect, isolated disc.}
\label{fig:mode_structure}
\end{figure}

\begin{align}
N_m &= \frac{1}{A_m} \int_0^R N(\rho) J_0(k_m \rho)^2 \rho \,d\rho \label{eq:Nmeqn},
\end{align}
with:
\begin{align}
A_m = \frac{R^2 J_1(k_m R)^2}{2}, \label{eq:Am}
\end{align}
and the radius-dependent demagnetizing factor defined as in Reference \cite{JosephandSchlomann1965}:
\begin{align}
N(\rho) &= \frac{R}{2} \int_0^\infty J_0(t \rho) J_1(t R)  (e^{-t z}+e^{-t (L - z)})\,dt . \label{eq:N_rho}
\end{align}
and the radius-dependent demagnetizing factor averaged over the thickness: \cite{JosephandSchlomann1965}
\begin{align}
N(\rho) &= -\frac{R}{L} \int_0^\infty J_0(t \rho) J_1(t R)  \frac{e^{-L t}-1}{t} \,dt \label{eqnNrho} .
\end{align}

\end{document}